\documentclass[acmlarge,screen]{acmart}
\usepackage{graphicx}
\usepackage{fontawesome}
\usepackage{float}
\usepackage{graphicx} 
\usepackage{subfigure}
\usepackage{algorithm}
\usepackage{algorithmic}
\usepackage{amsmath}

\AtBeginDocument{%
  \providecommand\BibTeX{{%
    \normalfont B\kern-0.5em{\scshape i\kern-0.25em b}\kern-0.8em\TeX}}}

\acmConference[ADVANCE '25]{International Workshop on ADVANCEs in ICT Infrastructures and Services}{June 23-25, 2025}{France}
\acmBooktitle{ADVANCE '25: Proceedings of the International Workshop on ADVANCEs in ICT Infrastructures and Services, June 23--25, 2025, Sophia-Antipolis, Nice, France} 

\begin{document}




\title{Towards Sustainability in 6G Network Slicing with Energy-Saving and Optimization Methods}

\author{Rodrigo Moreira}
\orcid{0000-0003-1310-9366}
\email{rodrigo@ufv.br}
\affiliation{%
  \institution{Universidade Federal de Viçosa (UFV)}
  \streetaddress{P.O. Box 1212}
  \city{Viçosa}
  \state{Minas Gerais}
  \country{Brazil}
  \postcode{}
}


\author{Tereza C. M. Carvalho}
\orcid{}
\email{terezacarvalho@usp.br}
\affiliation{%
  \institution{Universidade de São Paulo (USP)}
  \city{São Paulo}
  \country{Brazil}
}

\author{Flávio de Oliveira Silva}
\orcid{0000-0001-7051-7396}
\email{flavio@di.uminhop.pt}
\affiliation{%
  \institution{Universidade do Minho}
  \city{Braga}
  \country{Portugal}
}

\author{Nazim Agoulmine}
\orcid{}
\email{nagoulmine@gmail.com}
\affiliation{%
  \institution{Université Paris-Saclay - Évry}
  \city{Évry}
  \country{France}
}
\author{Joberto S. B. Martins}
\authornote{All authors contributed equally to this research.}
\orcid{0000-0003-1310-9366}
\affiliation{%
  \institution{Universidade Salvador (UNIFACS)}
  \streetaddress{Av. ACM 1133}
  \city{Salvador}
  \country{Brazil}}
\email{joberto.martins@gmail.com}

\renewcommand{\shortauthors}{Moreira and Martins}

\begin{abstract}

The 6G mobile network is the next evolutionary step after 5G, with a prediction of an explosive surge in mobile traffic. It provides ultra-low latency, higher data rates, high device density, and ubiquitous coverage, positively impacting services in various areas. Energy saving is a major concern for new systems in the telecommunications sector because all players are expected to reduce their carbon footprints to contribute to mitigating climate change. Network slicing is a fundamental enabler for 6G/5G mobile networks and various other new systems, such as the Internet of Things (IoT), Internet of Vehicles (IoV), and Industrial IoT (IIoT). However, energy-saving methods embedded in network slicing architectures are still a research gap. 
This paper discusses how to embed energy-saving methods in network-slicing architectures that are a fundamental enabler for nearly all new innovative systems being deployed worldwide. This paper's main contribution is a proposal to save energy in network slicing. That is achieved by deploying ML-native agents in NS architectures to dynamically orchestrate and optimize resources based on user demands. The SFI2 network slicing reference architecture is the concrete use case scenario in which contrastive learning improves energy saving for resource allocation.

\end{abstract}





\begin{teaserfigure}
   \centering
   \includegraphics[width=0.6\textwidth]{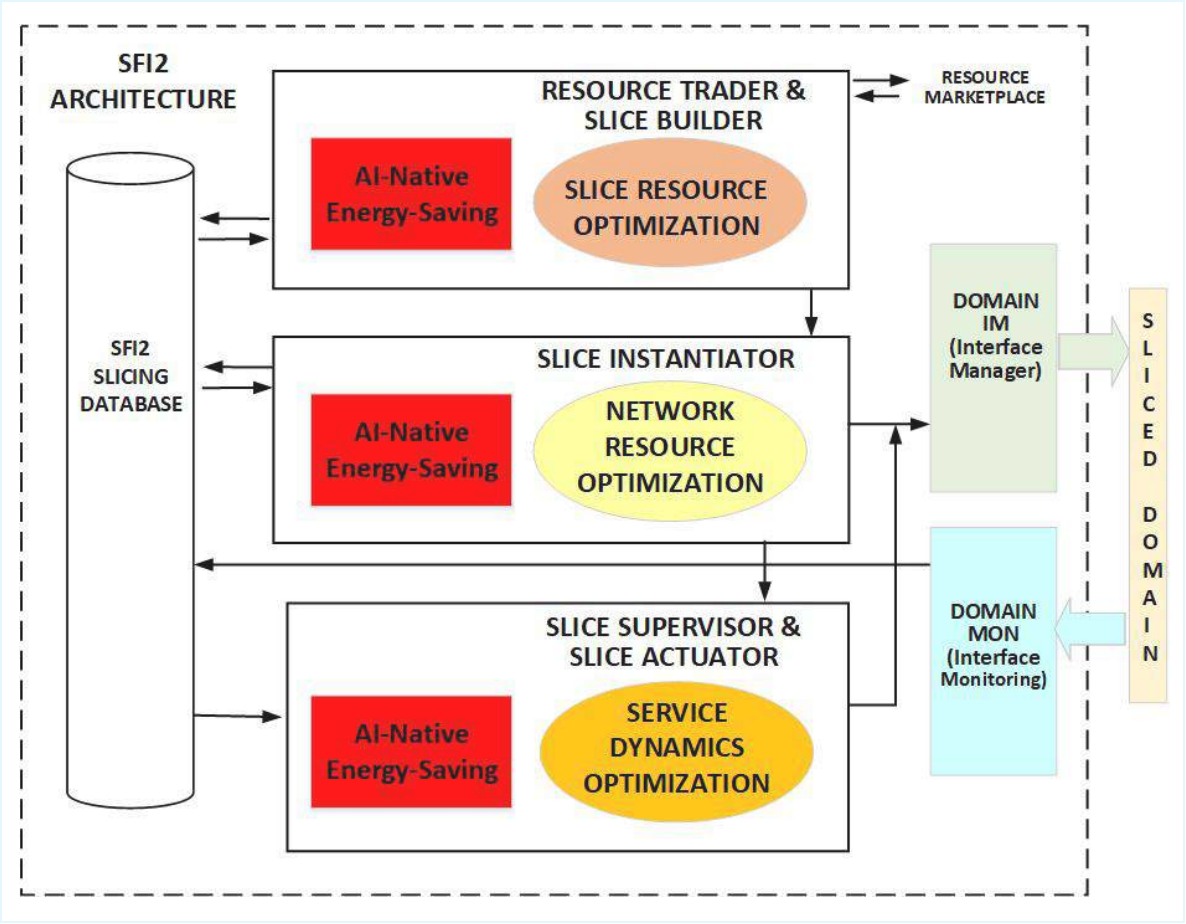}
  \caption{TEASER: 6G Energy-Saving Methods for Network Slicing Architectures}
  \Description{}
  \label{fig:teaser}
\end{teaserfigure}

\received{12 March 2025}
\received[revised]{30 April 2025}
\received[accepted]{15 05 25}


\maketitle

\section{Introduction}\label{sec:intro}

The 6G mobile network is the next step toward a new generation of high-speed mobile services, bringing new capabilities such as ultra-fast connection at terabit/second, high reliability, and ultra-low latency, support to new types of communication services such as immersive holographic and virtual reality communications, tactile communications, ubiquitous connectivity encompassing air, ground, and sea, self-sustainable network, etc. \cite{murroni_6genabling_2023}. The 6G will be globally disseminated and improve support for many productive sectors and verticals. Its global reach is to positively impact many future businesses as an enabler for highly innovative services, and its adoption is a matter of time \cite{okere_sixth_2025}.


Regarding the 6G telecommunications industry and climate change mitigation, most actions seek more sustainable solutions through carbon footprint reduction. In effect, carbon footprint reduction is one of the various approaches for climate change mitigation that has been extensively considered by various areas, including the telecommunications sector \cite{itu_measuring_2024}.

\begin{figure}[b]
  \includegraphics[width=0.6\textwidth]{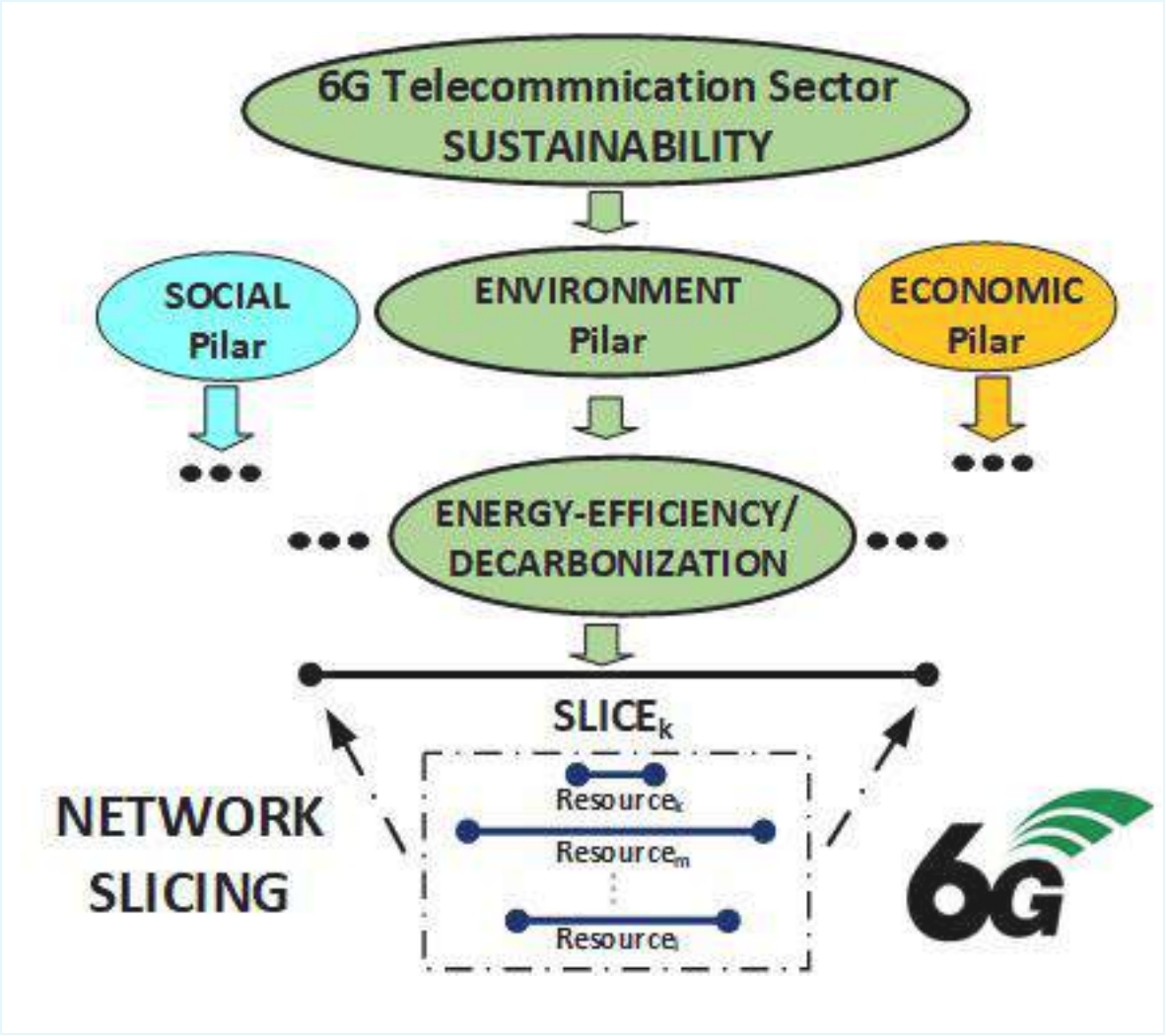}
  \caption{6G Telecommunication Sector - Sustainability.}
  \Description{6G Telecommunication Sector - Sustainability.}
  \label{fig:6G_Sustainability}
\end{figure}

In the 6G telecommunications sector, energy consumption optimization and reduction are two of the main contributors to reducing greenhouse gas emissions and, as such, contributing to climate change mitigation (Figure \ref{fig:6G_Sustainability}) \cite{taneja_power_2022}. As such, energy-efficient solutions are the main target and approach to promoting sustainable telecom solutions. In this regard, the telecommunications sector is aligned with climate change mitigation by defining and promoting a net zero\footnote{Zero carbon-dioxide emissions} target in which the telecommunications sector's carbon footprint is drastically reduced \cite{lou_towards_2024}. Complementary to the net zero targets, new key performance indicators (KPIs) are necessary to monitor the outcomes of carbon footprint solutions based on energy consumption reduction \cite{alliance_for_telecommunications_industry_solutions_6g_2023}.

However, in the current 6G research and development scenario, there is a gap in exploring opportunities for defining more ambitious new energy-saving solutions and energy-efficiency (EE) key performance indicators (KPIs).

This paper explores and addresses the gap in new energy-saving methods for 6G mobile services by focusing on the network slicing architectures and capabilities inherently necessary to support the 6G infrastructure. Network slicing is a fundamental component of 6G infrastructure that enables the multiplexing of virtualized and independent logical networks on the same physical network infrastructure. This work is, therefore, fundamental since promoting new energy-saving methods for slicing addresses the sustainability of 6G services.



Network slicing (NS) is, therefore, a cornerstone for 6G mobile network deployment because it supports virtual networks' planning, commissioning, configuration, operation, and management of all phases \cite{3gpp_3rd_2019}. By virtualizing physical and virtual resources like machines, communication links, switches, and radio access networks (RAN), among others, Network slicing allows service customization and infrastructure optimization that is crucial for 6G networks sustainable energy-saving mechanisms \cite{moreira_intelligent_2024a} \cite{subedi_network_2021} \cite{barakabitze_5g_2020}. 

Network Slicing is a flexible approach to provide the optimization capabilities that are necessary to address the highly dynamic and variable requirements imposed by future 6G mobile users \cite{hong_6g_2022}, future Vehicular Networks \cite{waheed_comprehensive_2022}, experimental networks \cite{martins_enhancing_2023}, industrial IIoT \cite{wu_survey_2022} and so on. Thanks to its capabilities to optimize and customize the network delivered to the end users, energy-saving should be addressed accordingly to reduce the CO2 footprint in these various areas.



This paper's objective is, therefore, to present the energy-efficient solutions that are envisioned in the context of the Slicing Future Internet Infrastructures (SFI2) reference architecture\footnote{https://sites.google.com/view/sfi2/home}. It benefits from its innovative characteristics as described in Martins et al. in \cite{donatti_survey_2023}.

This paper is organized as follows: The introduction section \ref{sec:intro} presents the vision of future telecommunications networks and highlights the need for energy-saving mechanisms for 6G mobile networks. Sections \ref{sec:6G_NS} and \ref{sec:EE_IN_SFI2} present how the sustainability goal is achieved using energy-saving mechanisms for 6G deployments that are highly dependent on network slicing as an enabler. Section \ref{sec:EE_IN_SFI2} complementary proposes an ML-embedded energy-saving solution in the network slicing architecture frame. Section \ref{sec:UseCase} presents a use case to save energy in network slicing architectures, having SFI2  architecture as the architectural reference. Finally, Section \ref{sec:Conclusion} closes the discussion with the final considerations.

\section{6G Slicing with Energy-Saving for Sustainability} 
\label{sec:6G_NS}





6G networks must provide enhanced speeds, global coverage, improved service capacity, high reliability, and ultra-low latency, to name a few stringent capabilities and requirements, while minimizing energy consumption. Sustainability through energy-efficient methods is a foundational element of 6G research and design, which implements new technologies and architectures with a sustainability mindset \cite{matinmikko-blue_multi-perspective_2024}.

Standardization institutions like 3GPP (3rd Generation Partnership Project), ITU-T (International Telecommunications Union - Telecommunications), and ETSI (European Telecommunications Standards Institute) have proposed NS architectures with various features and distinct target domains \cite{donatti_survey_2023}. Although this plurality of options, network slicing architectures have a common group of functional phases. The 3GPP network slicing standard and initiative defines the following  phases  \cite{3gpp_3rd_2019}:
\begin{itemize}
    \item The \textit{Preparation Phase} in which the user's slice requests are received, and the necessary resources are identified and localized in the target domain.
     
    \item The \textit{Commissioning Phase} in which the NS service provider makes choices and orchestrates among the available resources aiming to configure the requested slice.
    
    \item The \textit{Operation Phase} in which the deployed slice is operational and dynamic orchestration of resources may occur due to the user's time-varying demands and traffic fluctuation.
    
    \item The \textit{Decommissioning Phase} in which the allocated slice resources from single or multi-domains are liberated.
\end{itemize}

\begin{figure}[H]
  \includegraphics[width=0.6\textwidth]{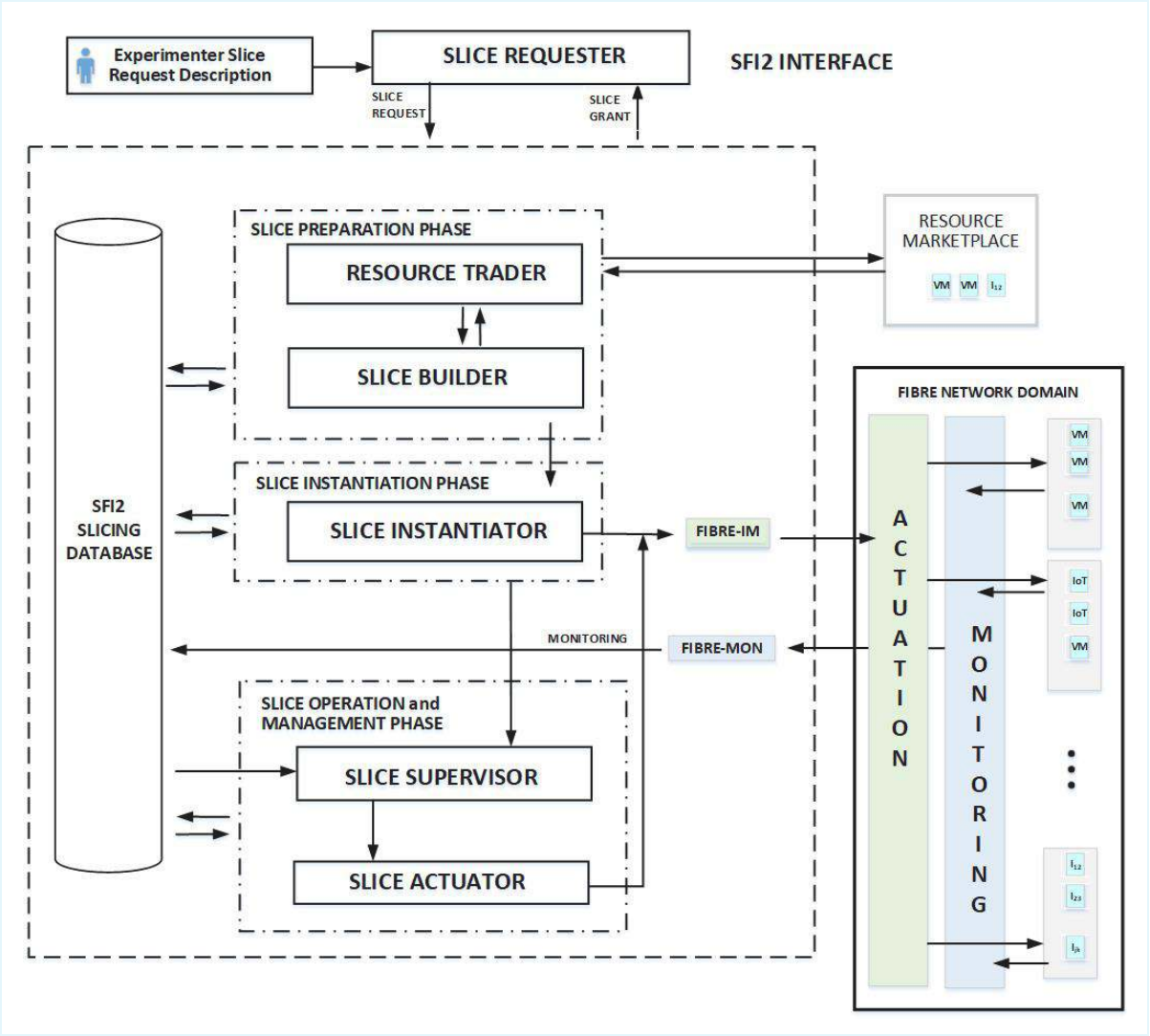}
  \caption{The Slicing Future Internet Infrastructures (SFI2) Network Slicing Reference Architecture \cite{martins_enhancing_2023}.}
  \Description{The Slicing Future Internet Infrastructures (SFI2) Network Slicing Reference Architecture}
  \label{fig:SFI2_Arch}
\end{figure}

The SFI2 project \cite{martins_enhancing_2023} defines the SFI2 network slicing reference architecture to create and manage multi-domain and multi-technology end-to-end slicing services. This architecture incorporates advanced network slicing concepts like ML-native optimizations, energy-efficient slicing, and slicing-tailored security functionalities \cite{moreira_enhancing_2023} \cite{moreira_intelligent_2025} as highlighted in Figure \ref{fig:SFI2_Arch}.

In the SFI2 architecture, the slice builder component builds the requested slice, considering the available resources in the domain marketplace, and optimizes their utilization in each built slice 
In sequence, the slice instantiation component deploys the configured slice in the target 
and optimizes the resources made available to the provider's network. 
At this phase, the optimization concerns 
the ensemble of resources available by the slicing service provider. Once the slice is deployed, the SFI2 slice supervisor component manages the slice's operation, allowing slice reconfiguration to cope with 
changes in the traffic, changes in user dynamic, or SLA (Service Level Agreement) tuning, among other possibilities \cite{martins_enhancing_2023}.


As previously highlighted, 6G infrastructure and services that telecommunication service providers will deploy will extensively use network slicing to virtualize and share resources. In this context, the main question is how NS architectures will contribute to and promote sustainability.

In this research paper, we approach this question from the perspective of energy-saving methods. In other words, to promote sustainability in 6G network slicing architectures, we aim to propose a novel solution to render the setup and operation of network slicing energy-saving aware. 


The idea is that energy consumption and RAN resource optimization are minimized during the preparation, deployment, and supervision phases of slice deployment, taking into account the users' demands.




\section{6G Resource Orchestration and Energy-Saving in SFI2 Network Slicing Architecture} \label{sec:EE_IN_SFI2}

The computer industry has multiple pathways to energy efficiency and sustainable solutions since the sustainable strategy adopted will differ among sectors and embrace various social, economic, and environmental aspects. This leads to potentially multiple deployed solutions 
across the industry.

\begin{figure}[H]
  \includegraphics[width=0.6\textwidth]{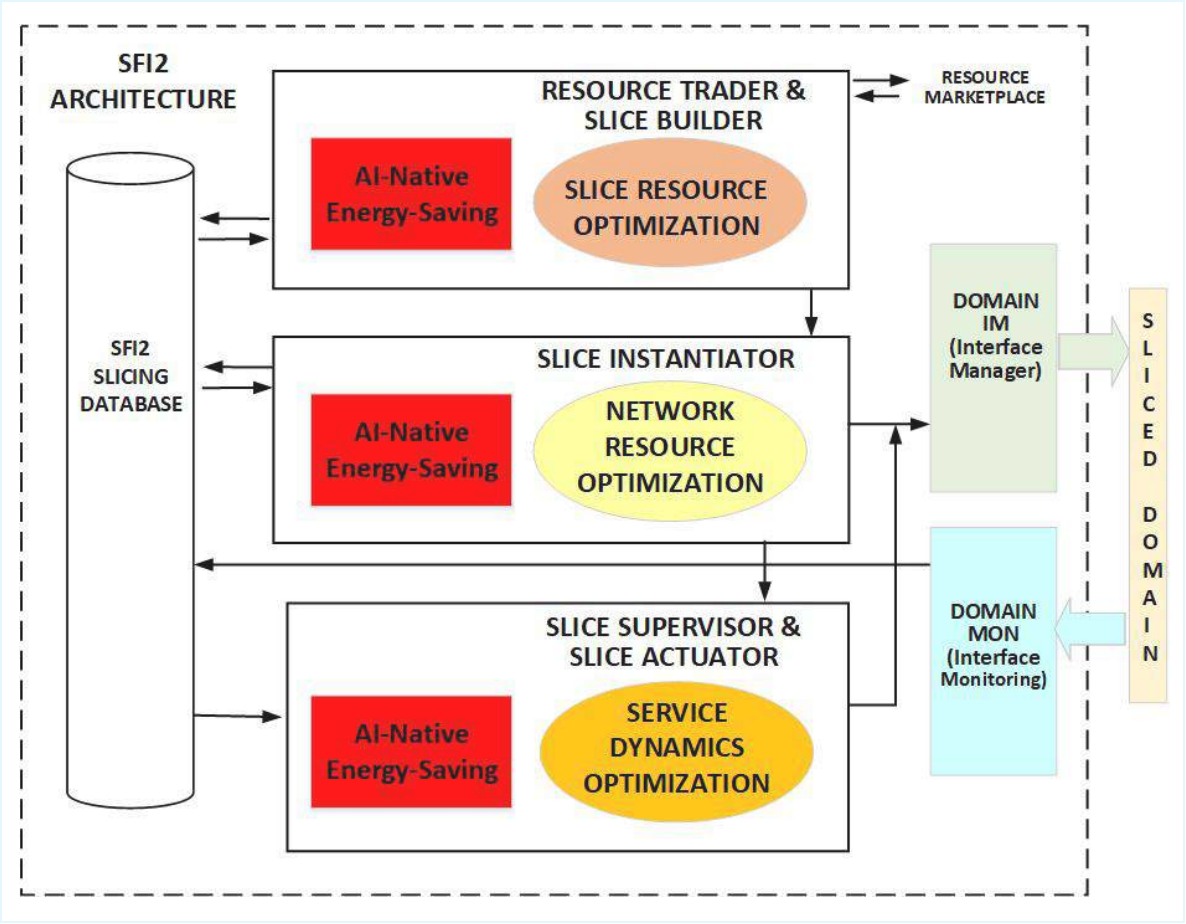}
  \caption{6G Resource Orchestration and Optimization towards Energy Saving in SFI2 Architecture.}
  \Description{6G Resource Orchestration and Optimization towards Energy Saving in SFI2 Architecture.}
  \label{fig:AI-Native-SFI2}
\end{figure}

In the context of the SFI2 network architecture, the 6G resource orchestration optimizations consider the following three optimization possibilities as illustrated in (Figure \ref{fig:AI-Native-SFI2}):
\begin{itemize}
    \item Slice Resource Optimization (SRO): occurs while resources are selected and deployed for a specific slice to provide the required service in the slicing-building process. The approach consists of optimizing resources considering a slice individually. 

    \item Network Resource Optimization (NRO): takes place during the instantiation phase of the network slicing process and concerns the set of slices provided by the service provider. In this case, the optimization process considers the set of slices active and the pool of resources available for the slice provider.

    \item Service Dynamic Optimization (SDO): This is executed during the slice operation and will deal with optimizing resources concerning the real-time services provided by the slice. This approach will consider a set of slices that may comprise either all active slices or any set of them, depending on the actual services being deployed by the provider.

\end{itemize}

\section{Proposed Solution: Network Slicing Energy Saving using Contrastive Learning}\label{sec:UseCase}




In this paper, we propose using Contrastive Learning (CL)  \cite{jaiswal_survey_2021} to optimize network slicing and save energy by improving resource allocation, load balancing, and decision-making processes.

The idea is to use contrastive learning models to compare traffic patterns and demand forecasts for different slices and identify similarities and differences in traffic behaviors. Based on that, system resources like computing power, bandwidth, and others can be allocated efficiently.


The proposed solution aims to introduce ML agents in the SFI2 architecture to monitor the energy consumption of the processes allocated to instantiate and supervise slices.

The system model considers the SFI 2 architecture's Service Dynamic Optimization (SRO) and Network Resource Optimization (NRO), along with embedded ML agents (using CL) to monitor the allocated process's energy consumption and provide forecasting information to SRO and NRO to achieve their objectives, taking sustainability into account. 



\subsection{Contrastive Learning for Detecting Anomalies in Unlabeled Energy Consumption Time Series}



Contrastive learning is a robust method for learning data representations without labels. It aims to bring similar data points closer together in the representation space while pushing dissimilar points apart. This approach is effective for detecting anomalies in time series data, which often lack labeled examples. Given a set of time series data \( \{x_i\}_{i=1}^{N} \), where \( x_i \) represents an instance, the objective is to learn a representation \( f(x_i) \) that clusters normal patterns and separates anomalies. The contrastive loss function is according to Equation~\ref{eq:contrastive1}.

\begin{equation}
\label{eq:contrastive1}
\mathcal{L}_{\text{contrastive}} = \sum_{i=1}^{N} \sum_{j=1}^{N} \ell(f(x_i), f(x_j), y_{ij})
\end{equation}

where \( \ell(f(x_i), f(x_j), y_{ij}) \) is the loss between instances \( x_i \) and \( x_j \), and \( y_{ij} \) indicates similarity (1) or dissimilarity (0). A common form of the loss is according to Equation~\ref{eq:contrastive2}.

\begin{equation}
\label{eq:contrastive2}
\ell(f(x_i), f(x_j), y_{ij}) = y_{ij} \| f(x_i) - f(x_j) \|^2 + (1 - y_{ij}) \max(0, m - \| f(x_i) - f(x_j) \|)^2
\end{equation}

Here, \( m \) is a margin parameter, \( \| \cdot \| \) is the Euclidean norm, and \( f(x) \) is the learned representation. The function \( f \) is typically implemented using deep neural networks like CNNs or RNNs, which capture temporal dependencies. To detect anomalies, measure the distance between representations. Instances outside the normal cluster are considered anomalies. For a new instance \( x' \), the anomaly score \( s(x') \) is according to Equation~\ref{eq:contrastive3}.

\begin{equation}
\label{eq:contrastive3}
s(x') = \min_{i=1}^{N} \| f(x') - f(x_i) \|
\end{equation}

If \( s(x') \) exceeds a threshold, \( x' \) is an anomaly. We defined empirically a threshold of 0.5.

 For our use case, we employed a datacenter energy-consumption dataset containing 15 features~\cite{Estrada2022}. Data were collected over a period of 120 days, with a sampling frequency of one measurement per hour. Among these features, we selected the five best ones by the correlation method to be used as an object of anomaly analysis, namely, Voltage, Current, Power, Frequency, and Energy, in the order of relevance measured by the method. We conducted a preliminary analysis to validate how anomaly detection methods with unlabeled data are capable of identifying anomalies with the least amount of noise possible.

\begin{figure}[H]
  \includegraphics[width=0.8\textwidth]{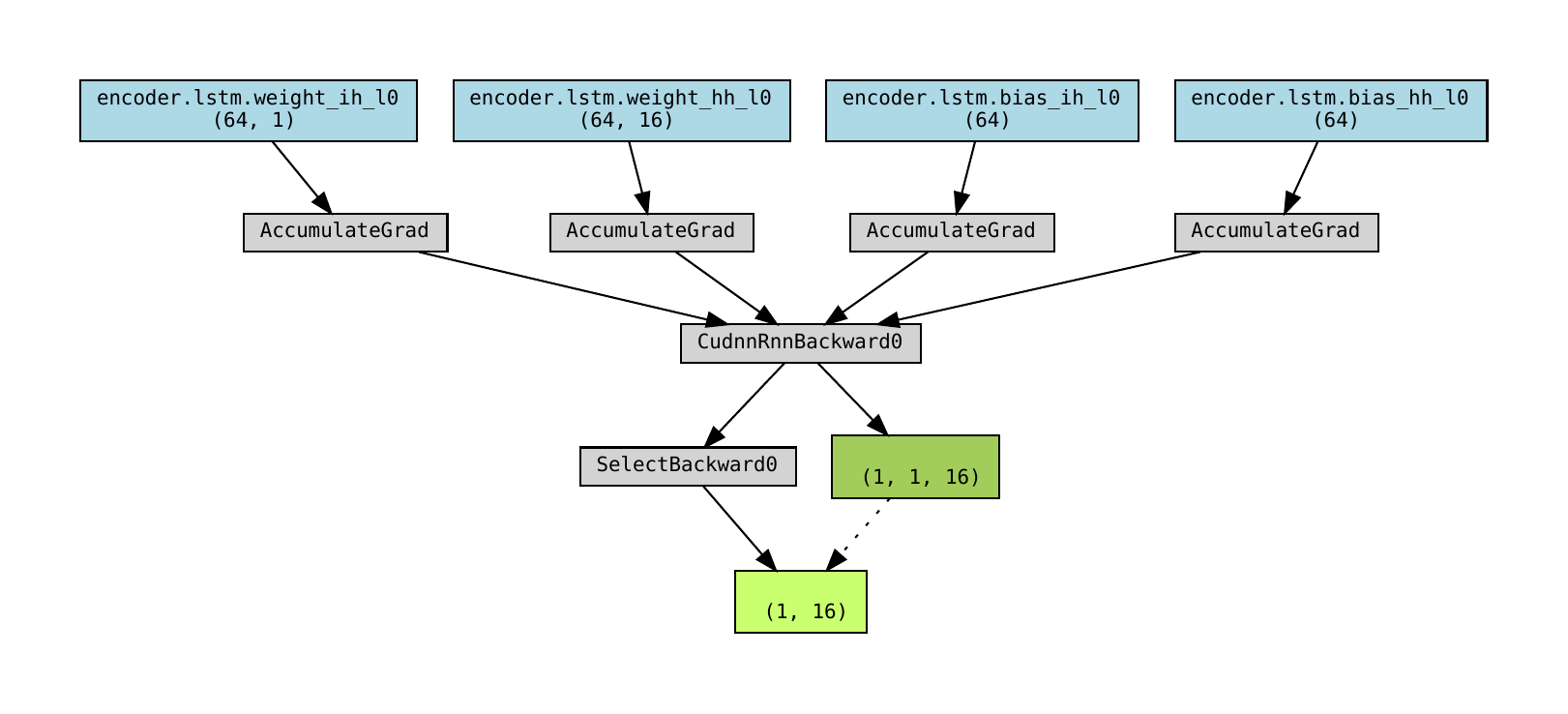}
  \caption{A LSTM model for Contrastive Learning.}
  \Description{A LSTM model for Contrastive Learning.}
  \label{fig:model_architecture}
\end{figure}

 Our Contrastive Learning method utilizes a Time Series Encoder with an LSTM layer to capture the temporal dependencies in time-series data using 16 hidden units. The encoded output was fed into a contrastive model to enhance feature discrimination by contrasting positive and negative pairs. An Adam optimizer was employed to adjust the parameters efficiently. A visualization of the model architecture is shown in Fig. ~\ref{fig:model_architecture}, which illustrates the sequential connections and operations, highlighting gradient flow and backpropagation. 

Detecting anomalies using contrastive learning involves comparing pairs of data points to determine their similarity. The process can be summarized in the Algorithm~\ref{alg:anomaly}.

\begin{algorithm}
\caption{Detect Anomalies}
\begin{algorithmic}[1]
\STATE \textbf{Input:} Column $column$, Model $model$, DataFrame $df$, Threshold $threshold$
\STATE \textbf{Output:} Vector of Anomalies

\STATE $model.eval()$
\STATE $(X_i, X_j) \gets \text{create\_pairs}(column, df)$
\STATE $X_i, X_j \gets X_i.to(device), X_j.to(device)$
\STATE $(h_i, h_j) \gets model(X_i.unsqueeze(-1), X_j.unsqueeze(-1))$
\STATE $distances \gets \text{torch.norm}(h_i - h_j, \text{dim}=1)$
\STATE $anomalies \gets distances > threshold$
\RETURN $anomalies$
\end{algorithmic}
\label{alg:anomaly}
\end{algorithm}

 Algorithm~\ref{alg:anomaly} begins by receiving the input parameters: a column of data (column), a machine learning model (model), a DataFrame (df), and an anomaly detection threshold (threshold). The model was set to the evaluation mode (model.eval()) to ensure proper functioning during inference. Data pairs ($X\_i$, $X\_j$) were created from the specified column in the DataFrame. These pairs were then transferred to an appropriate computational device. The model processes pairs to generate their respective embeddings ($h\_i$, $h\_j$). The distance between these embeddings was calculated using the Euclidean norm. Anomalies were identified by comparing these distances to a predefined threshold, with distances exceeding the threshold indicating potential anomalies. The algorithm then returns a vector of the identified anomalies.

In the context of unlabeled time series data, k-means clustering and skewness can be effectively used to identify anomalies. K-means clustering involves partitioning the time series data into \( k \) clusters by minimizing the variance within each cluster. The formula for updating the centroid of a cluster is given by Equation~\ref{eq:contrastive4}

\begin{equation}
\label{eq:contrastive4}
\mu_j = \frac{1}{|C_j|} \sum_{x_i \in C_j} x_i 
\end{equation}

where \( \mu_j \) is the centroid of cluster \( j \), \( C_j \) is the set of points assigned to cluster \( j \), and \( x_i \) represents the data points. Anomalous points can be identified as those that are far from their assigned cluster centroids. Skewness, a measure of the asymmetry of the data distribution, can further enhance anomaly detection. The skewness \( \gamma \) is calculated according to Equation~\ref{eq:contrastive5}

\begin{equation}
\label{eq:contrastive5}
\gamma = \frac{n}{(n-1)(n-2)} \sum \left(\frac{x_i - \bar{x}}{s}\right)^3 
\end{equation}

where \( n \) is the number of data points, \( \bar{x} \) is the mean, \( s \) is the standard deviation, and \( x_i \) represents each data point. High skewness values can indicate the presence of outliers or anomalies in the data. By combining the clustering results from k-means with skewness analysis, one can robustly detect and interpret anomalies in unlabeled time series data.

 We report these results in Table~\ref{table:anomalies}, where it can be observed that Contrastive Learning captured anomalies with the lowest incidence of noise.

\begin{table}[H]
\centering
\begin{tabular}{|c|c|c|c|}
\hline
\textbf{Feature} & \textbf{\begin{tabular}[c]{@{}c@{}}Contrastive Learning\\ with LSTM\end{tabular}} & \textbf{K-Means} & \textbf{Skewness} \\
\hline
Voltage & $2093.2 \pm 874$ & $112972 \pm 26295$ & $69821 \pm 0$ \\
Current & $8115 \pm 4271$ & $87731 \pm 34729$ & $80380 \pm 0$ \\
Power & $ 65073 \pm 35727 $ & $262117 \pm 105794 $ & $ 50560 \pm 0 $ \\
Frequency & $18597 \pm 8021$ & $332186 \pm 12262$ & $89431 \pm 0$ \\
Energy & $26 \pm 11$ & $494502 \pm 17353$ & $2241 \pm 0$ \\
\hline
\end{tabular}
\caption{Detected Anomalies by Method and Feature}
\label{table:anomalies}
\end{table}

\begin{figure}[H]
    \centering
    \subfigure[Current.]{\includegraphics[width=0.30\textwidth]{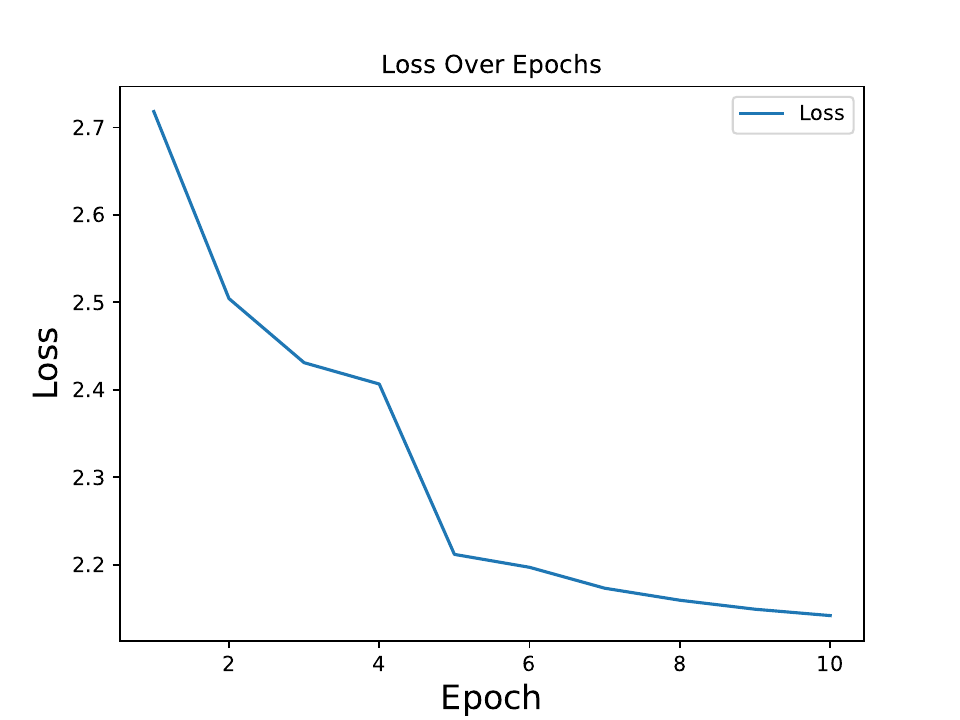}}
    \subfigure[Energy.]{\includegraphics[width=0.30\textwidth]{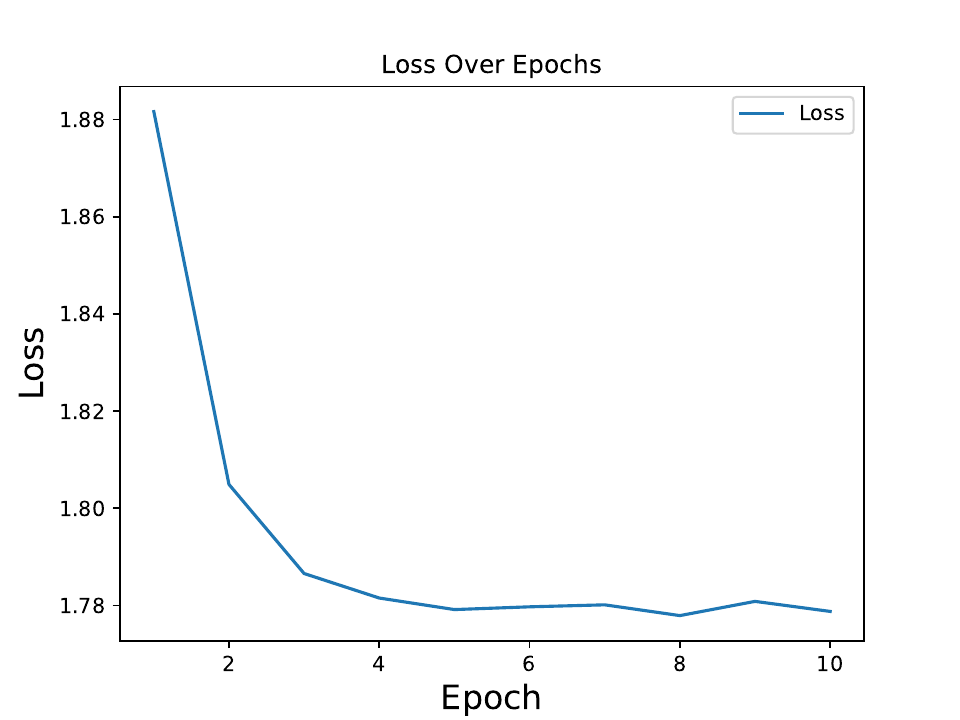}}
    \subfigure[Frequency.]{\includegraphics[width=0.30\textwidth]{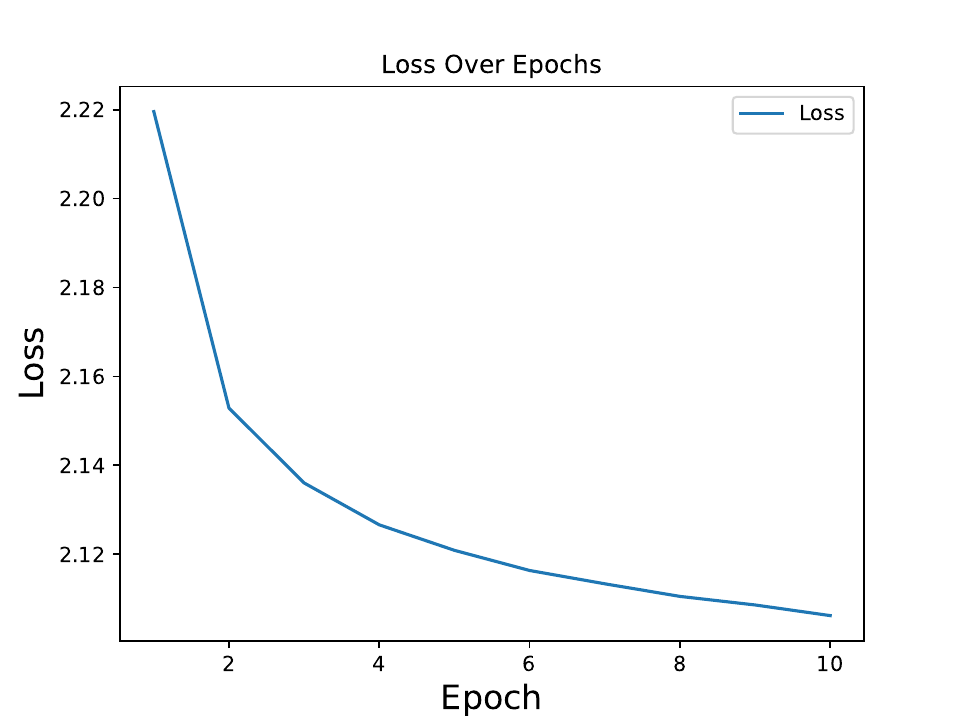}}
    \subfigure[Power.]{\includegraphics[width=0.30\textwidth]{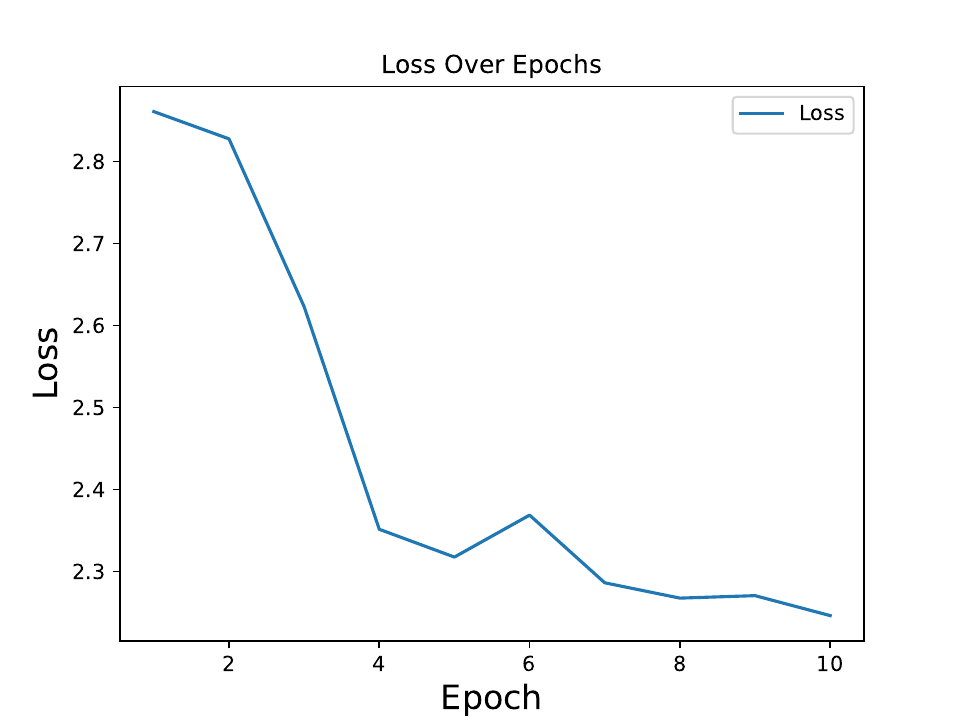}}
    \subfigure[Voltage.]{\includegraphics[width=0.30\textwidth]{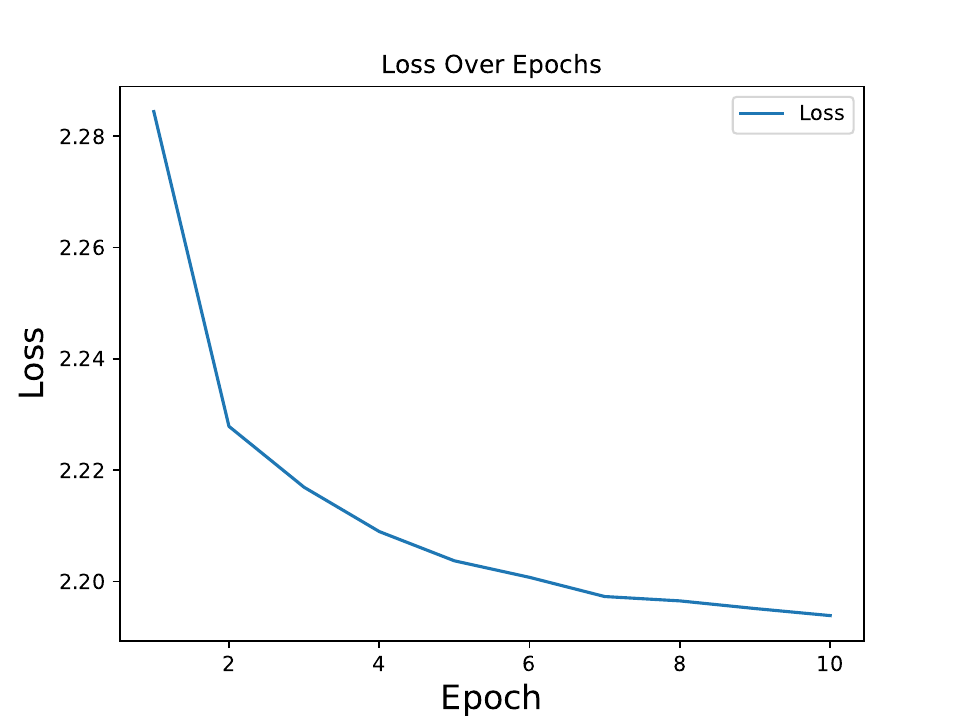}}
    \caption{Losses over training time.}
    \Description{Losses over training time.}
    \label{fig:losses}
\end{figure}

With respect to contrastive learning with LSTM, we confirm in Fig.~\ref{fig:losses}, there was learning in the model without overfitting in the task of detecting anomalies in the five features considered. We empirically used a learning rate of $10^{-3}$ and Adam optimizer.

\begin{figure}[htbp]
  \includegraphics[width=0.5\textwidth]{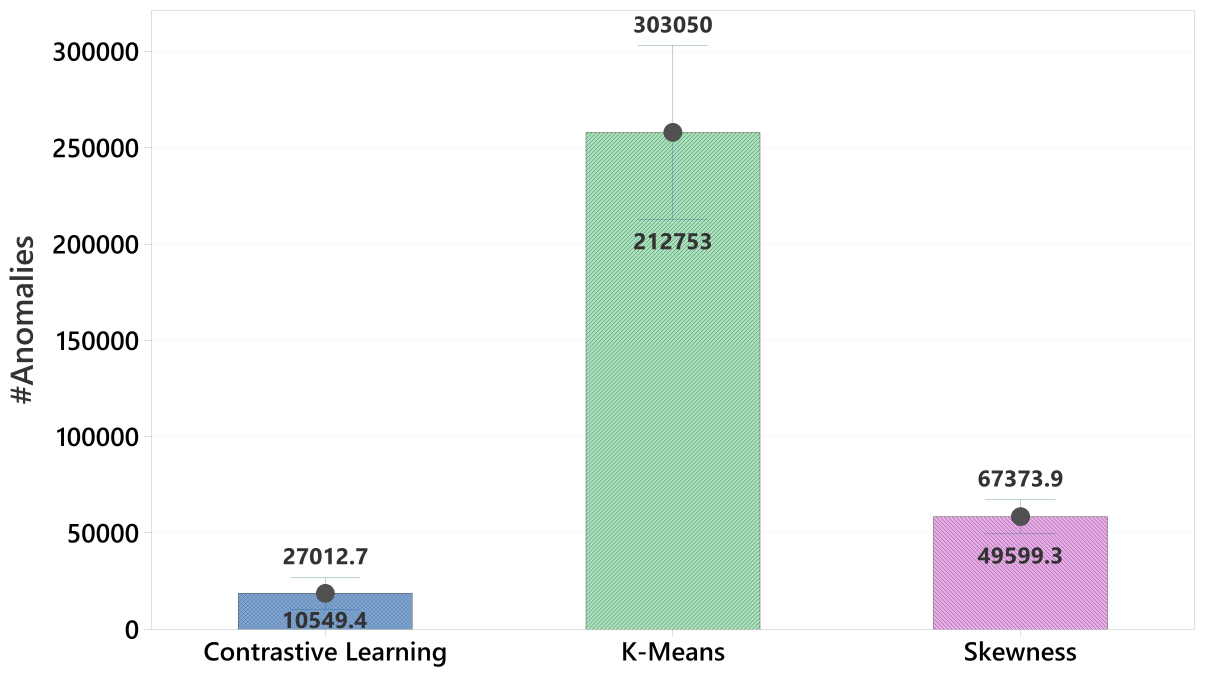}
  \caption{Amount of noised anomalies detected using this method.}
  \Description{Amount of noised anomalies detected using this method.}
  \label{fig:anomalies_detected_by_method}
\end{figure}

Finally, we measured the quantity and anomalies detected by each method, namely Contrastive Learning, K-means, and Skewness. We report in Fig.~\ref{fig:anomalies_detected_by_method} and verify that contrastive learning focuses on distinguishing meaningful differences by learning a representation that groups similar patterns while separating dissimilar ones. This method reduces noise sensitivity and captures more genuine structural variations in the time series data, leading to visually coherent and plausible anomaly detections. Although the lower number of detected anomalies might seem conservative, the model's improved alignment with visual expectations suggests it may more accurately represent true anomalies. This outcome underscores contrastive learning’s potential in balancing anomaly detection sensitivity with robustness to noise, presenting a promising direction for handling unlabeled time series anomaly detection tasks.

K-means, for instance, is sensitive to outliers since its cost function minimizes the distance between data points and cluster centers, which can lead to clustering anomalies alongside noisy data and obscure precise anomaly patterns. Skewness, which measures data distribution asymmetry, can reveal significant deviations in the shape of time series but is also influenced by extreme variations, often amplifying noise presence. Together, these effects result in anomaly detection that frequently flags noise as relevant deviations, raising false positive rates. This behavior underscores the potential need for more robust methods to distinguish noise patterns from genuine anomalies, suggesting an exploration of preprocessing techniques or algorithms that are less sensitive to non-representative data fluctuations.

\section{Final Considerations} \label{sec:Conclusion}

Energy-efficient and energy-saving strategies are crucial in the contemporary era of network architectures. In the state of the art, there are directions towards a network for more than connectivity; while guaranteeing stringent metrics, it is imperative to consider energy efficiency in the early stages of network architecture design. This paper elucidates how modern network slicing architectures can benefit from machine learning-based energy-aware slicing control plane approaches. This study incorporates results from embedding contrastive learning in the SFI2 Slicing Architecture to identify energy consumption anomalies in data centers where slicing and applications are deployed. The findings indicate that contrastive learning is less susceptible to noise, enabling AI algorithms to accurately capture energy demands throughout the slicing lifecycle. Future work will aim to integrate this approach with real-time energy probes to accurately estimate and inform control plane slicing for modern network architectures.

\begin{acks}

The authors thank the FAPESP MCTIC/CGI cooperation agreement under the thematic research project 2018/23097-3 - Slicing Future Internet Infrastructures (SFI2),  Brazilian National Council for Scientific and Technological Development (CNPq), grant \# 421944/2021-8, FAPEMIG (Grant APQ00923-24) and the ANIMA Institute for scholarship support 2024/2025.

\end{acks}


\bibliographystyle{unsrt}
\bibliography{ADVANCE}









\end{document}